\documentclass[aps,prd,twocolumn,nofootinbib,superscriptaddress,showpacs,amsmath,amssymb]{revtex4}
\usepackage{graphicx,epsf}
\usepackage[]{latexsym}
\newcommand{\ve}{\varepsilon}
\newcommand{\be}{\begin{eqnarray}}
\newcommand{\ee}{\end{eqnarray}}
\newcommand{\bea}{\begin{eqnarray}}
\newcommand{\eea}{\end{eqnarray}}

\def\comment#1{}

\newcommand{\lp}{\ell_{\rm p}}
\newcommand{\mpl}{m_{\rm p}}
\newcommand{\gn}{G_{\rm N}}
\newcommand{\rh}{r_{\rm H}}

\newcommand{\ep}{\mathcal{E}_{\rm p}}

\usepackage{color}

\definecolor{darkred}{rgb}{.8,0,0}

\definecolor{darkblue}{rgb}{0,0,.7}

\definecolor{darkgreen}{rgb}{0,.7,0}

\begin{document}

%
%
\title{GUP parameter from quantum corrections to the Newtonian potential}

%
\author{Fabio~Scardigli}
\email{fabio@phys.ntu.edu.tw}
\affiliation{Dipartimento di Matematica, Politecnico di Milano, Piazza L.da Vinci 32, 20133 Milano, Italy}
\affiliation{Department of Applied Mathematics, University of Waterloo, Ontario N2L 3G1, Canada}
\affiliation{Yukawa Institute for Theoretical Physics,
Kyoto University, Kyoto 606-8502, Japan}
\author{Gaetano~Lambiase}
\email{lambiase@sa.infn.it}
\affiliation{Dipartimento di Fisica "E.R. Caianiello", Universita' di Salerno, I-84084 Fisciano (Sa), Italy \&\\
INFN - Gruppo Collegato di Salerno, Italy}
\author{Elias~C.~Vagenas}
\email{elias.vagenas@ku.edu.kw}
\affiliation{Theoretical Physics Group, Department of Physics,
Kuwait University, P.O. Box 5969, Safat 13060, Kuwait}
%
%
%
%
%
%
\begin{abstract}
\par\noindent
We propose a technique to compute the deformation parameter of the generalized uncertainty principle
by using the leading quantum corrections to the Newtonian potential.  We just assume General 
Relativity as theory of Gravitation, and the thermal nature of the GUP corrections to the Hawking spectrum. 
With these minimal assumptions our calculation gives, to first order, a specific numerical result. The physical 
meaning of this value is discussed, and compared with the previously obtained bounds on the generalized uncertainty 
principle deformation parameter.
\end{abstract}
%

%
\maketitle
%
%
%
%
\section{Introduction}
%
%
%
\par\noindent
Research on generalizations of the uncertainty principle (GUP) of quantum mechanics has
nowadays a long history~\cite{GUPearly}.
One of the main lines of investigation focuses on understanding how the Heisenberg
Uncertainty Principle (HUP) should be modified once gravity is taken into account.
Given the pivotal role played by gravitation in these arguments, it is not
surprising that the most relevant modifications to the HUP have been proposed in
string theory, loop quantum gravity, deformed special relativity, and studies of
black hole physics~\cite{VenezGrossMende,MM,kempf,FS,Adler2,SC}.
\par
As it is well know, the dimensionless deforming parameter of the GUP,
henceforth denoted by $\beta$, is not (in principle) fixed by the theory, although it is
generally assumed to be of order one (this happens, in particular, in some models of string theory,
 see for instance Ref.~\cite{VenezGrossMende}).
\par
There have been many studies that aim at setting bounds on $\beta$, for instance Refs.  \cite{brau},
\cite{vagenas}, \cite{Nozari}.
In these works, a specific (in general, non linear) representation of the operators in the deformed
fundamental commutator is utilized~\footnote{We shall work with $c=k_B=1$, but explicitly
show the Newton constant $\gn$ and Planck constant $\hbar$.
We also recall that the Planck length is defined as $\lp^2=\gn\,\hbar/c^3$,
the Planck energy as $\ep\,\lp = \hbar\, c /2$, and the Planck mass as
$\mpl=\ep/c^2$, so that $\gn=\lp/2\,\mpl$ and $\hbar=2\,\lp\,\mpl$.}
\be
\left[\hat{X},\hat{P}\right] = i\,\hbar\left(1 + \beta\, \frac{\hat{P}^2}{\mpl^2}\right)
\label{[1]}
\ee
in order to compute corrections to quantum mechanical quantities, such as
energy shifts in the spectrum of the hydrogen atom, or to the Lamb shift,
the Landau levels, Scanning Tunneling Microscope, charmonium levels, etc.
The bounds so obtained on $\beta$
are quite stringent, ranging from $\beta < 10^{21}$ to $\beta < 10^{50}$.
\par
A further group of bounds can be found in Refs. \cite{LNChang} and \cite{Nozari2},
where a deformation of classical Newtonian mechanics
is introduced by modifying the standard Poisson brackets in a way that resembles
the quantum commutator
\be
\left[\hat{x},\hat{p}\right]
=
i\,\hbar\left(1 + \beta_0\, \hat{p}^2\right)
\,
\Rightarrow
\,
\{X,P\} = \left(1 + \beta_0\, P^2\right)\,
\ee
where $\beta_0=\beta/\mpl^2$.
However, in the limit $\beta \to 0$, Ref.~\cite{LNChang}
recovers {\em only} the Newtonian mechanics but {\em not} GR, and GR corrections must
be added as an extra structure.
Clearly, the physical relevance of this approach and the bound that follows
for $\beta$ remains therefore questionable.
\par
Finally, in Refs. \cite{ghosh} and \cite{pramanik}, the authors consider the gravitational
interaction when evaluating bounds on $\beta$.
They use a covariant formalism which firstly is defined in Minkowski space, with the metric
$\eta_{\mu\nu}={\rm diag}(1,-1,-1,-1)$, which can be easily generalized to curved space-times
via the standard procedure $\eta_{\mu\nu} \to g_{\mu\nu}$.
In addition, these papers, as the previous ones, start from a deformation of classical Poisson brackets,
although posited in covariant form.
From the deformed covariant Poisson brackets, they obtain interesting consequences,
like a $\beta$-deformed geodesic equation, which leads to a violation of the Equivalence Principle.
This formalism remains covariant when $\beta \to 0$
and  it reproduces the standard GR results in the limit $\beta \to 0$ (unlike papers as Ref. \cite{LNChang}).
\par
Among the papers which consider the gravitational interaction when evaluating bounds on $\beta$, it is
Ref.~\cite{SC2}. This approach differs from the previous ones because
Poisson brackets and classical Newtonian mechanics remain untouched. Additionally,
GR and standard quantum mechanics are recovered, when $\beta \to 0$. Therefore, the
Equivalence Principle is  preserved, and the equation of motion of
a test particle is still given by the standard geodesic equation.
The bounds on $\beta$ proposed by papers which take into account gravity range from
$\beta<10^{19}$, for those papers admitting a violation of equivalence principle, to $\beta < 10^{69}$
for the papers preserving the aforesaid principle.
\par
In the present paper, we exhibit a computation of the value of $\beta$ obtained by comparing
two different low energy (first order in $\hbar$) corrections for the expression of the Hawking temperature.
The first is due to the GUP, and therefore involves $\beta$. The second correction, instead, is obtained by
including the deformation of the metric due to quantum corrections to the Newtonian potential.
Then we demand the two corrections to be equal (at the first order), and this yields a specific numerical
value for $\beta$. It results to be of order of unity, in agreement with the general belief and with some 
 particular models of string theory.
%
%
%
%
%
%
%
%
%
\section{GUP-deformed Hawking temperature}
%
%
%
%
\par\noindent
One of the most common forms of deformation of the HUP
(as well as the form of GUP that we are going to study in this paper) is
\be
\Delta x\, \Delta p
&\geq&
\frac{\hbar}{2}\left(1 \ + \ \beta\,\frac{4\,\lp^2}{\hbar^2}\,\Delta p\,^2\right) \nonumber \\
&=&
\frac{\hbar}{2}\left[ 1 + \beta \left(\frac{\Delta p}{\mpl} \right)^2\right]
\label{gup}
\ee
which, for mirror-symmetric states (with $\langle \hat{p} \rangle^2 = 0$),
can be equivalently written in terms of commutators as
\be
[\hat{x},\hat{p}]
=
i\hbar \left[1 + \beta \left(\frac{\hat{p}}{\mpl} \right)^2 \right]
\ee
since $\Delta x\, \Delta p \geq (1/2) |\langle [\hat{x},\hat{p}] \rangle|$.
\par
As is well known from the argument of the Heisenberg microscope~\cite{Heisenberg},
the size $\delta x$ of the smallest detail of an object, theoretically detectable
with a beam of photons of energy $E$, is roughly given by
\be
\delta x
\simeq
\frac{\hbar}{2\, E}
\label{HS}
\ee
since larger and larger energies are required to explore smaller and smaller details.
From the uncertainty relation~\eqref{gup}, we see that the GUP version of the standard
Heisenberg formula~\eqref{HS} is
\be
\delta x
\simeq
\frac{\hbar}{2\, E}
+ 2\,\beta\,\lp^2\, \frac{E}{\hbar}
\label{He}
\ee
which relates the (average) wavelength of a photon to its energy $E$
\footnote{Here, the standard dispersion relation $E=p\,c$ is assumed.}.
Conversely, using  relation~(\ref{He}), one can compute the energy
$E$ of a photon with a given (average) wavelength $\lambda \simeq \delta x$. 
To compute the thermal GUP corrections to the Hawking spectrum, we follow the arguments of 
Refs.~\cite{FS9506,ACSantiago,CavagliaD,CDM03,Susskind,nouicer,Glimpses}, and
we consider  an ensemble of unpolarized photons of Hawking radiation
just outside the event horizon of a Schwarzschild black hole.
From a geometrical point of view,
it is easy to see that the position uncertainty of such photons is of
the order of the unmodified Schwarzschild radius, i.e., $\rh=2\, G\,M$.
An equivalent argument comes from considering the average wavelength of
the Hawking radiation, which is of the order of the geometrical size
of the hole. We can estimate the uncertainty in photon position to be
$\delta x \simeq 2\,\mu\,\rh $, where the proportionality constant
$\mu$ is of order unity and will be fixed soon.
According to the equipartition principle, the average energy
$E$ of unpolarized photons of the Hawking radiation is simply
related with their temperature by $E = T$.
Inserting the aforesaid expressions for the uncertainty in the photon position
and for the average energy into  formula~(\ref{He}), we obtain
\be
{4\,\mu\,G\, M} \simeq \frac{\hbar}{2\,T}+ 2\,\beta\,G\,T~.
\label{37}
\ee
\par\noindent
In order to fix $\mu$, we consider the semiclassical limit
$\beta \to 0$, and require formula (\ref{37}) to predict the standard semiclassical
Hawking temperature, namely  $T(\beta\to 0)=T_{\rm H}$,
\be
T_{\rm H} = \frac{\hbar}{8\pi\, G\,M}~.
\label{Hw0}
\ee
This fixes $\mu = \pi$, thus we have
\be
M=\frac{\hbar}{8\pi\, G\, T}
+\beta\, \frac{T}{2\pi}~.
\label{MT}
\ee
This is the mass-temperature relation predicted by the GUP for a Schwarzschild
black hole. Of course this relation can be easily inverted, to get
\be
T = \frac{\pi}{\beta}\left(M-\sqrt{M^2-\frac{\beta}{\pi^2}\,\mpl^2} \right)~.
\label{TM}
\ee
%
%
%
However, since the term proportional to $\beta$ is small,
especially for solar mass black holes with $M\gg\mpl$,
we can expand in powers of $\beta$, namely
\be
T = \frac{\hbar}{8\pi GM} \left(1 + \frac{\beta\,\mpl^2}{4\pi^2\,M^2} + \dots \right)~.
\label{Tg}
\ee
and it is evident that to zero order in $\beta$, we recover the usual Hawking formula~(\ref{Hw0}).
\par\noindent
Once again we stress that we are assuming that the correction induced by the GUP has
a thermal character, and, therefore, it can be cast in the form of a shift of the Hawking temperature. Of course, there
are also different approaches, where the corrections do not respect the exact thermality of the spectrum, and thus need
not be reducible to a simple shift of the temperature. An example is the corpuscular
model of a black hole of Ref.~\cite{dvali}. In this model, the emission is expected to gain a
correction of order $1/N$, where $N \sim (M/\mpl)^2$ is the number of constituents, and it
becomes important when the mass $M$ approaches the Planck mass.
%
%
%
%
%
%
\section{Temperature from a deformed Schwarzschild metric}
%
\subsection{Leading quantum correction to the Newtonian potential}
%
%
%
%
\par\noindent
After early results by Duff  \cite{Duff:1974ud}, the leading quantum correction to the Newtonian potential
has been computed by Donoghue,  by assuming General Relativity as fundamental theory of Gravity. 
In a series of beautiful papers (see for instance Ref.~\cite{Dono})
he reformulated General Relativity as an effective field theory, and, in particular, he considered two heavy bodies
close to rest. The leading quantum correction derived from this model shows a long-distance \emph{quantum} effect.
More recently,  Donoghue and other authors found that the gravitational interaction between
the two objects can be described by the potential energy  \cite{Dono2}
\be
U(r)=-\frac{G M m}{r}\left(1 + \frac{3 G(M+m)}{rc^2} + \frac{41}{10\pi}\frac{\ell_P^2}{r^2}\right) .
\label{johnD}
\ee
The first correction term does not contain any power of $\hbar$, so it is a classical effect, due to the
non-linear nature of General Relativity. However, the second correction term, i.e., the last term of  (\ref{johnD}),
is a true quantum effect, linear in $\hbar$. The potential generated by the mass $M$ reads
\be
V(r)=-\frac{GM}{r}\left(1 + \frac{3GM}{r}(1+\frac{m}{M}) +  \frac{41}{10\pi}\frac{\ell_P^2}{r^2}\right) .
\label{DP}
\ee
%
%
%
%
%
%
\subsection{Effective potential from the metric}
%
%
%
%
%
\par\noindent
Now we  consider the effective potential produced by a metric of the very general class
\be
ds^2 = F(r)dt^2 - g_{ik}(x_1,x_2,x_3) dx^i dx^k
\label{ik}
\ee
where $r=|\bold x|=(x_1^2 + x_2^2 + x_3^2)^{1/2}$, and $x_1,x_2,x_3$ are the standard Cartesian coordinates.
Particular cases of the metric (\ref{ik})  is  the Schwarzschild metric, in the standard form
\be
ds^2 = \left(1-\frac{2GM}{r}\right)dt^2 - \left(1-\frac{2GM}{r}\right)^{-1}dr^2 - r^2d\Omega^2 \nonumber
\ee
as well as in harmonic coordinates
\be
ds^2 &=& \left(\frac{R-GM}{R+GM}\right)dt^2 - \left(\frac{R+GM}{R-GM}\right)dR^2 \nonumber \\
&-& (R+GM)^2 d\Omega^2 \,.\nonumber
\ee
with  $R=r-GM$.
\par\noindent
It can be easily seen \footnote{More details can be found in Ref.  \cite{Weinberg72}.}
that any general metric of the form
\be
ds^2 = F(r)dt^2 - F(r)^{-1}dr^2 - C(r)d\Omega^2
\label{gm}
\ee
can be put in the form (\ref{ik}).
In fact, Eq. (\ref{gm}) is equivalent to
\begin{eqnarray}
ds^2 &=& F(r)dt^2 - \left(F(r)^{-1} - \frac{C(r)}{r^2}\right)\frac{1}{r^2}({\bf x} \cdot d{\bf x})^2 \nonumber \\
 &-& \frac{C(r)}{r^2}d{\bf x}^2\,. \nonumber
\end{eqnarray}
Once the metric is in the form of (\ref{ik}), in Cartesian coordinates, then, with well known procedures  \cite{Weinberg72},
it is easy to show that the effective Newtonian potential \footnote{The effective Newtonian potential is produced
by the metric given  in (\ref{ik}) for  a point particle which  moves slowly, in a stationary  and weak gravitational
field, i.e., quasi-Minkowskian far from the source, $r \to \infty$.} is of the form
\be
V(r) \ \simeq \ \frac{1}{2} \ (F(r)-1)
\ee
or,  equivalently,
\be
F(r) \ \simeq \ 1 \ + \ 2 \ V(r)~.
\ee
%
%
%
%
\subsection{Metric mimicking the quantum corrected Newtonian potential}
%
%
%
%
%
\par\noindent
At this point, we can  write down the metric which is able to mimic the quantum corrected
Newtonian potential proposed by Donoghue. Recalling (\ref{DP}), we have
\be
&&F(r) \ \simeq \ 1 + 2 V(r) \ =  \nonumber \\
&&1 - \frac{2GM}{r} - \frac{6\, G^2 M^2}{r^2}\left(1+\frac{m}{M}\right) -
\frac{41}{5\pi}\frac{G^3 M^3}{r^3}\left(\frac{\ell_P}{GM}\right)^{2}.\nonumber
\ee
Let us now define
\be
\epsilon(r) = - \frac{6\, G^2 M^2}{r^2}\left(1+\frac{m}{M}\right) -
\frac{41}{5\pi}\frac{G^3 M^3}{r^3}\left(\frac{\ell_P}{GM}\right)^{2}\hspace{-1.5ex}.~
\label{phi}
\ee
Therefore, $F(r)$ will now be of the form
\be
F(r) = 1 - \frac{2GM}{r} + \epsilon(r)
\label{DM}
\ee
and it is evident that when $r$ is large, then  $|\epsilon(r)| \ll 2GM/r$.
%
%
%
%
%
%
%
%
%

\subsection{Computing $\beta$}
%
%
%
%
\par\noindent
We can legitimately wonder what kind of (deformed) metric would predict a Hawking temperature
like the one inferred from the GUP in relation~(\ref{Tg}), for a given $\beta$.
Since we are interested only in small corrections to the Hawking formula,
we can consider a deformation of the Schwarzschild metric of the following kind
\footnote{Recently, it was argued that in the special case in which $\epsilon(r) \sim 1/r^{2}$,
the specific metric (\ref{GDSCH}) could have some drawbacks  in the context
of GUP formalism \cite{Ali:2015zua}. However, none of those drawbacks appear here and, thus,
there is no problem to employ (\ref{GDSCH}) in our present study.}
\be
F(r) = 1 - \frac{2 \,G\, M}{r} + \epsilon(r)
\label{GDSCH}
\ee
where $\epsilon(r)$ is an arbitrary, small, smooth function of $r$.
We note that the deformation (\ref{GDSCH}) makes sense when $|\epsilon(r)| \ll GM/r$.
We can also introduce a regulatory small parameter $\ve$ and, thus,  we can write $\epsilon(r) \equiv \ve\phi(r) $.
At the end of the calculation, $\ve$ can go to unity.
Of course, we look for the lowest order correction in the dimensionless parameter $\ve$.
The horizon's equation, i.e., $F(r)=0$, now reads
\be
r - 2 \,G\, M + \ve\, r\, \phi(r) = 0~.
\label{eqr}
\ee
Such equations  can be solved, in a first approximation in $\ve$, as follows.
First, we formulate (\ref{eqr}) in a general form
\be
x = a + \ve f(x)~.
\label{eqx}
\ee
It is obvious that if $\ve$ is set equal to zero, then the solution will be $x_0=a$.  If $\ve$  is slightly different from zero,
then we can try a test solution of the form $x_0 = a + \eta(\ve)$
where $\eta(\ve) \to 0$ for $\ve \to 0$. Substituting the aforesaid test solution in (\ref{eqx}), we get $x_0 = a + \ve f(x_0)$
which means $\eta = \ve f(a+\eta)$.
To first order in $\eta$, we have $\eta = \ve[f(a) + f'(a)\eta]$ from which we obtain
$\eta = \frac{\ve f(a)}{1-\ve f'(a)}$.
Therefore, to first order in $\ve$, the general solution of (\ref{eqx})  reads  $x_0 \ = \ a + \frac{\ve f(a)}{1-\ve f'(a)}$.
Applying this formula to (\ref{eqr}), we get the solution
\be
r_H \ = \ a - \frac{\ve\, a\, \phi(a)}{1 \ + \ \ve\,[\phi(a) \ + \ a\,\phi'(a)]}
\label{solr}
\ee
where $a = 2 G M$.
%
%
%
%
%
%
\par\noindent
The Hawking temperature is given by
\be
T =  \frac{\hbar}{4\pi}F'(r_H)~.
\label{hawtemp}
\ee
From Eq. (\ref{GDSCH}), one gets
\be
F'(r) = \frac{a}{r^2} \ + \ \ve\,\phi'(r)\,.
\ee
It is useful to write the solution (\ref{solr}) in the compact form $r_H = a(1-\lambda)$
where $\lambda = \frac{\ve\, \phi(a)}{1 \ + \ \ve\,[\phi(a) \ + \ a\,\phi'(a)]}$  and,
therefore, $\lambda \sim \ve$, $|\lambda| \ll 1$.
Then
\be
F'(r_H) = \frac{1}{a(1-\lambda)^2} \ + \ \ve \phi'[a(1-\lambda)]\,.
\ee
Therefore, the deformed Hawking temperature reads
\be
T &=& \hbar \frac{F'(r_H)}{4\pi} = \frac{\hbar}{4\pi a} \left\{1 + \ve \left[2\phi(a) + a\phi'(a)\right] \right. \nonumber \\
&+& \left. \ve^2 \phi(a) \left[\phi(a) - 2a\phi'(a) - a^2 \phi''(a) \right] + \dots \right\}~.
\label{teps}
\ee
It is noteworthy that the only function $\phi(r)$ that  annihilates the first-order in $\ve$  temperature
correction term is the solution of the differential equation $2\phi(r) + r\phi'(r) \ = \ 0$,
namely $\phi(r) = A / r^2$, where $A$ is an arbitrary constant.
In particular, for the function $\phi(r) = G^2 M^2 / r^2$, the coefficient of $\ve$ in (\ref{teps}) is zero,
and the coefficient of $\ve^2$ is $-1/16$.
It is also interesting to investigate what kind of function will eliminate the second-order in $\ve$ correction term.
This function will be the solution of the differential equation $r^2 \phi''(r) + 2r\phi'(r) - \phi(r) \ = \ 0$ %
which is an Euler equation.
Its characteristic equation is of the form $\lambda^2 + \lambda -1 = 0$
with roots $\lambda_1 = \frac{-1 - \sqrt{5}}{2}$ and  $ \lambda_2 = \frac{-1 + \sqrt{5}}{2}$~.
So, the functions which remove  the  $\ve^2$-correction term are
$\phi_1(r) = r^{-|\lambda_1|}$ and  $\phi_2(r) = r^{\lambda_2}$~.
\par\noindent
We are now in the position to compute the temperature generated by the metric (\ref{DM}), by simply
employing  (\ref{teps}). Therefore,  the metric-deformed Hawking temperature  is of the form
\footnote{Notice that from (\ref{phi}) we have $\epsilon(r) \sim 1$ for $r \sim a$,
so this would seem to spoil the expansion (\ref{tesp2}) when $r \sim a$. On the contrary, we can always
imagine to first expand the temperature $T(r)=\hbar F'(r)/4\pi$ for $r \gg a$, when $\epsilon(r)$ is small.
Then, the term in $1/r^2$ disappears from the expansion of $T(r)$
 because of the condition $2\phi(r) + r\phi'(r) \ = \ 0$. Finally, we take the limit $r \to a$, and this yields (\ref{tesp2}).}
\be
T &=&\frac{\hbar}{4\pi a} \left\{1 + \left[2\epsilon(a) + a\epsilon'(a)\right] + \dots \right\}
\label{tesp2}
\ee
while the GUP-deformed Hawking temperature reads
\be
T = \frac{\hbar}{8\pi GM} \left(1 + \frac{\beta\,\mpl^2}{4\pi^2\,M^2} + \dots \right)~.
\ee
By comparing the two respective first-order correction terms in the two aforesaid expansions, we obtain
\be
\beta \ = \  \frac{4\pi^2 M^2}{\mpl^2}\, \left[2\epsilon(a) + a\epsilon'(a)\right]~.
\label{beta1}
\ee
Using now expression (\ref{phi}) for $\epsilon(r)$, we get
\be
2\epsilon(r) + r\epsilon'(r) =  \frac{B}{r^3}
\ee
with  $B=\frac{41 \, \, G^3 M^3}{5 \pi}\left(\frac{\ell_P}{GM}\right)^2 $.
Therefore,
\be
2\epsilon(a) + a\epsilon'(a) = \frac{B}{8 G^3 M^3}
\ee
and using (\ref{beta1}), the parameter $\beta$ will get the value
\be
\beta = \frac{4\pi^2 M^2}{\mpl^2}\,\frac{41}{40 \pi}\left(\frac{\ell_P}{GM}\right)^2
= \frac{82\pi}{5}~.
\ee
%
%
%
%
%
%
%
\section{Conclusions}
%
%
%
%
%
\par\noindent
In this work we have computed the value of the deformation parameter $\beta$ of the GUP.
We obtain this result by computing in two different ways the Hawking temperature for a Schwarzschild black hole.
\par
The first way consists in using the GUP (in place of the standard HUP) to compute the Hawking formula.
In this way  we get an expression of the temperature containing a correction term
depending on $\beta$, i.e.,  the GUP-deformed Hawking temperature (\ref{Tg}).
\par
The second way involves the consideration of the quantum correction to the Newtonian potential, computed years ago
by Donoghue and others. The corrections to the Newtonian potential imply naturally a quantum correction to the
Schwarzschild metric. Therefore, the Hawking temperature computed through this quantum corrected Schwarzschild
metric result to get corrections in respect to the standard Hawking expression,
i.e., the metric-deformed Hawking temperature (\ref{tesp2}).
\par
The request that the first-order corrections of  the two different expressions of Hawking temperature
must coincide,  fixes unambiguously the numerical value of $\beta$ to be $82\pi/5$.
\par\noindent
Finally, a couple of comments are in order here.
First, this numerical value is of order one, as expected from several string
theory models, and from versions of GUP derived through gedanken experiments.  
In particular, this is the first time, to our knowledge, that a specific value is obtained for $\beta$ 
by starting from the minimal assumptions we made.
Second, as we know, in the last years much research has focused on the
experimental bounds of the \emph{size\/} of $\beta$, and several
experiments have been proposed to test GUPs in the laboratory. In fact, it has been shown that one does 
not need to reach the Planck energy scale to test GUP corrections.
Among the more elaborated proposals, where conditions can be created in a lab, are those of the groups of
Refs.~\cite{brukcerd,cerd,bonaldi}. However, it is also worth of note that  the best bounds on $\beta$ presented 
in the literature are still by far much larger than the value computed here. This could require, presumably, a big 
leap in the experimental designs and techniques in order to search this region for the parameter $\beta$.
%
%
%
%
%
%
%
%
%
%
%
%

%
%
%
\end{document}